\documentclass[sigconf]{aamas} 

\usepackage{xfrac}
\usepackage[ruled,vlined]{algorithm2e}
\usepackage{algorithmic}

\setcopyright{ifaamas}
\acmConference[AAMAS '23]{The 14th Workshop on Optimization and Learning in Multiagent Systems (OptLearnMAS-23)
AT the 22nd International Conference
on Autonomous Agents and Multiagent Systems (AAMAS 2023)}{May 29 -- June 2, 2023}
{London, United Kingdom}{A.~Ricci, W.~Yeoh, N.~Agmon, B.~An (eds.)}
\copyrightyear{2023}
\acmYear{2023}
\acmDOI{}
\acmPrice{}
\acmISBN{}

\acmSubmissionID{???}

\title[AAMAS-2023 Formatting Instructions]{Coordinating Fully-Cooperative Agents Using Hierarchical Learning Anticipation}

\author{Ariyan Bighashdel}
\affiliation{
  \institution{Eindhoven University of Technology}
  \city{Eindhoven}
  \country{The Netherlands}}
\email{a.bighashdel@tue.nl}

\author{Daan de Geus}
\affiliation{
  \institution{Eindhoven University of Technology}
  \city{Eindhoven}
  \country{The Netherlands}}
\email{d.c.d.geus@tue.nl}

\author{Pavol Jancura}
\affiliation{
  \institution{Eindhoven University of Technology}
  \city{Eindhoven}
  \country{The Netherlands}}
\email{p.jancura@tue.nl}

\author{Gijs Dubbelman}
\affiliation{
  \institution{Eindhoven University of Technology}
  \city{Eindhoven}
  \country{The Netherlands}}
\email{g.dubbelman@tue.nl}

\begin{abstract}
Learning anticipation is a reasoning paradigm in multi-agent reinforcement learning, where agents, during learning, consider the anticipated learning of other agents. There has been substantial research into the role of learning anticipation in improving cooperation among self-interested agents in general-sum games. Two primary examples are Learning with Opponent-Learning Awareness (LOLA), which anticipates and shapes the opponent's learning process to ensure cooperation among self-interested agents in various games such as iterated prisoner's dilemma, and Look-Ahead (LA), which uses learning anticipation to guarantee convergence in games with cyclic behaviors. So far, the effectiveness of applying learning anticipation to fully-cooperative games has not been explored. In this study, we aim to research the influence of learning anticipation on coordination among common-interested agents. We first illustrate that both LOLA and LA, when applied to fully-cooperative games, degrade coordination among agents, causing worst-case outcomes. Subsequently, to overcome this miscoordination behavior, we propose Hierarchical Learning Anticipation (HLA), where agents anticipate the learning of other agents in a hierarchical fashion. Specifically, HLA assigns agents to several hierarchy levels to properly regulate their reasonings. Our theoretical and empirical findings confirm that HLA can significantly improve coordination among common-interested agents in fully-cooperative normal-form games. With HLA, to the best of our knowledge, we are the first to unlock the benefits of learning anticipation for fully-cooperative games.
\end{abstract}


\newcommand{\BibTeX}{\rm B\kern-.05em{\sc i\kern-.025em b}\kern-.08em\TeX}

\begin{document}


\pagestyle{fancy}
\fancyhead{}


\maketitle 


\section{Introduction}






One of the key characteristics of learning in multi-agent systems is the non-stationary environment. As a result, agents should continuously interact with each other and adapt their strategies accordingly. However, in various game settings, these interactions commonly lead to worst-case outcomes for all agents \citep{foerster2018learning}. A prominent example is the case of general-sum games, particularly the Iterated Prisoner’s Dilemma (IPD), where self-interested agents typically converge to defect-defect, which is the worst result globally. Recently, the learning anticipation paradigm, where agents take into account the anticipated learning of other agents, has been broadly employed to avoid such catastrophic outcomes \cite{foerster2018learning,letcher2018stable,zhang2010multi}. 

For instance, the Learning with Opponent-Learning Awareness (LOLA) method \citep{foerster2018learning} has proven to be successful in the IPD game. Specifically, in LOLA, agents anticipate and shape the learning step of others to enforce cooperation among themselves. This so-called \textit{opponent shaping} in LOLA leads to a tit-for-tat strategy in IPD (starting out cooperating and otherwise mirroring the opponent’s last move), which achieves mutual cooperation when chosen by both players \citep{axelrod1981evolution}. Subsequent works include: a) Higher-order LOLA (HOLA) \cite{foerster2018learning} that assumes non-na\"ive opponents, b) Look-Ahead (LA)\cite{zhang2010multi,letcher2018stable} that discards opponent shaping to guarantee convergence in games with cyclic behaviors (e.g., matching pennies), c) Stable Opponent Shaping (SOS) \cite{letcher2018stable} that guarantees convergence to the fixed points (i.e., points with zero gradients) in differentiable games, and d) Consistent LOLA (COLA) \cite{willi2022cola} which provides a closed-form solution of infinite-order LOLA.

Nevertheless, all of these methods are proposed to improve cooperation in games with self-interested agents. So far, no works have explored how these methods perform when agents are fully-cooperative, i.e., common-interested. Current investigations on fully-cooperative games are limited to the study of Letcher et al. \citep{letcher2018stable}, proving the convergence and non-convergence of LOLA to stable and unstable fixed points, respectively. However, we demonstrate that in fully-cooperative normal-form games with unstable fixed points, although the agents in both LOLA and LA do not converge to the unstable fixed points, they are subject to miscoordination and, consequently, worse overall rewards.

Considering the above, the key research goal of this work is to research the influence of learning anticipation on coordination among fully-cooperative agents. We believe that as multiple agents should interact and cooperate to achieve a common goal, learning anticipation has the potential to improve coordination among agents, resulting in better overall rewards.

To accomplish our goal, we first theoretically prove that in a two-agent two-action coordination game \citep{claus1998dynamics}, both LOLA and LA have the tendency to lead to miscoordination among common-interested agents, causing a worse outcome. To solve this miscoordination problem and improve the applicability of learning anticipation to fully-cooperative games, we then propose Hierarchical Learning Anticipation (HLA), a new learning method explicitly developed for improving coordination in games with common-interested agents. Specifically, HLA assigns all agents to hierarchy levels, which define the reasoning orders of agents. We theoretically prove that HLA can avoid miscoordination in the aforementioned coordination game. Furthermore, we empirically show that in a two-agent three-action coordination game \citep{claus1998dynamics}, HLA, as opposed to LOLA and LA, significantly improves coordination among agents, leading to better overall rewards. Finally, we discuss the shortcomings of our study and provide future research directions.



\section{Background}
Our work assumes a multi-agent task that is commonly described as a Markov Game (MG) \citep{littman1994markov}. An MG can be defined as a tuple $(\mathcal{N},\mathcal{S},\{\mathcal{A}_i\}_{i \in \mathcal{N}},\{\mathcal{R}_i\}_{i \in \mathcal{N}},\mathcal{T},\rho,\gamma)$, where $\mathcal{N}$ is the set of agents ($|\mathcal{N}|=n$), $\mathcal{S}$ is the set of states, and $\mathcal{A}_i$ is the set of possible actions for agent $i \in \mathcal{N}$. Agent $i$ chooses its action $a_i \in \mathcal{A}_i$ through the policy network $\pi_{\theta_i}:\mathcal{S} \times \mathcal{A}_i \rightarrow [0,1]$ parameterized by $\theta_i$ conditioning on the given state $s \in \mathcal{S}$. Given the actions of all agents, each agent $i$ obtains a reward $r_i$ according to its reward function $\mathcal{R}_i:\mathcal{S}\times \mathcal{A}_1 \times ... \times \mathcal{A}_{n} \rightarrow \mathbb{R}$. Given an initial state, the next state is produced according to the state transition function $\mathcal{T}:\mathcal{S} \times \mathcal{A}_1 \times ... \times \mathcal{A}_{n} \times \mathcal{S} \rightarrow [0,1]$. Given an episode $\tau$ of horizon $T$, the discounted return for each agent $i$ at time step $t \leq T$ is defined by $G^t_i(\tau)=\sum_{l=t}^T \gamma^{l-t} r_i$ where $\gamma$ is a predefined discount factor. The expected return given the agents’ policy parameters approximates the state value function for each agent  $V_i(s,\theta_1,...,\theta_{n})=\mathbb{E}[G_i^t(\tau|s^t=s)]$. Each agent $i$ aims to maximize the expected return given the distribution of the initial state $\rho(s)$, denoted by the performance objective $J_i = \mathbb{E}_{\rho(s)} V_i(s,\theta_1,...,\theta_{n})$. A \textit{na\"ive agent} updates its policy parameters in the direction of the objective's gradient 
\begin{equation}
\label{eq:naive}
\begin{split}
    \nabla_{\theta_i} J_i = \mathbb{E}_{\rho(s)}\nabla_{\theta_i} V_i(s,\theta_1,...,\theta_{n}).
\end{split}
\end{equation}

\textbf{Learning With Opponent-Learning Awareness (LOLA)}. Unlike na\"ive agents, LOLA agents modify their learning objectives by differentiating through the anticipated learning steps of the opponents \citep{foerster2018learning}. Given $n=2$ for simplicity, a first-order LOLA agent assumes a na\"ive opponent and uses policy parameter anticipation to optimize $V_1^{\text{LOLA}}(s,\theta_1,\theta_2+\Delta\theta_2)$ where $\Delta\theta_2 = \mathbb{E}_{\rho(s)}  \eta\nabla_{\theta_2}V_2(s,\theta_1,\theta_2)$ and $\eta\in \mathbb{R}^+$ is the prediction length. Using first-order Taylor expansion and by differentiating with respect to $\theta_1$, the gradient adjustment for the first LOLA agent~\citep{foerster2018learning}~is given by
\begin{equation}
\label{eq:lola}
\begin{split}
    \nabla_{\theta_1}V_1^{\text{LOLA}}(s,\theta_1,\theta_2+\Delta\theta_2) \approx &\nabla_{\theta_1}V_1 + (\nabla_{\theta_2\theta_1}V_1)^\intercal \Delta\theta_2\\
    &+ \underbrace{ (\nabla_{\theta_1} \Delta\theta_2)^\intercal \nabla_{\theta_2}V_1}_{\text{shaping}},
\end{split}
\end{equation}
where $V_1 = V_1(s,\theta_1,\theta_2)$. The rightmost term in the LOLA update allows for active shaping of the opponent's learning. This term has been proven effective in enforcing cooperation in various games with self-interested agents, including IPD \citep{foerster2018learning,foerster2018dice}. The LOLA update can be further extended to non-na\"ive opponents, resulting in HOLA agents \citep{foerster2018learning,willi2022cola}.

\textbf{Look Ahead (LA).} LA agents assume that the opponents' learning steps cannot be influenced, i.e., cannot be shaped \citep{zhang2010multi,letcher2018stable}. In other words, agent one assumes that the prediction step, $\Delta{\theta}_2$, is independent of the current optimization, i.e., $\nabla_{\theta_1} \Delta{\theta}_2 = 0$. Therefore, the shaping term disappears, and the gradient adjustment for the first-order LA agent will be
\begin{equation}
\label{eq:la}
\begin{split}
    \nabla_{\theta_1}V_1^{\text{LA}}(s,\theta_1,{\theta}_2+\perp\Delta{\theta}_2) \approx \nabla_{\theta_1}V_1 + (\nabla_{{\theta}_2\theta_1}V_1)^\intercal \Delta{\theta}_2,
\end{split}
\end{equation}
where $\perp$ prevents gradient flowing from $\Delta{\theta}_2$ upon differentiation. 

The benefits of LOLA and LA have been frequently shown throughout the literature in games with self-interested agents \cite{zhang2010multi,foerster2016learning,letcher2018stable}. In the next section, we analyze the effectiveness of LOLA and LA in fully-cooperative games with common-interested agents, i.e., $\mathcal{R}_i=\mathcal{R}_j \;\forall i,j \in \mathcal{N}$ and, consequently, $V_i=V_j \;\forall i,j \in \mathcal{N}$.


\section{Miscoordination analysis in fully-cooperative games}

To investigate the influence of LOLA and LA on coordination among common-interested agents, we consider a two-agent two-action coordination game \citep{claus1998dynamics} with an added miscoordination penalty. The game is defined by a common reward matrix
\begin{equation}
\begin{split}
    \mathcal{R}_1=\mathcal{R}_2=\begin{bmatrix}\alpha & k\\k & \alpha\end{bmatrix},
\end{split}
\end{equation}
where $\alpha >0 $ is the coordination reward and $k \leq 0$ is the miscoordination penalty. We further define $g=\alpha-k>0$ as the miscoordination regret. The agents are parameterized by $\theta_1\in[0,1]$ and $\theta_2\in[0,1]$, denoting the probability of choosing the first action by agents one and two, respectively. With a joint strategy $(\theta_1,\theta_2)$, the common value function of the game is
\begin{equation}
\begin{split}
    V_1(\theta_1,\theta_2)=V_2(\theta_1,\theta_2)=2g\theta_1 \theta_2 - g(\theta_1+\theta2) + \alpha,
\end{split}
\end{equation}
The game has two equilibrium points, i.e., $(\theta_1=0,\theta_2=0)$ and $(\theta_1=1,\theta_2=1)$, where each agent receives a reward of $\alpha$. Furthermore, two miscoordination points of the games are $(\theta_1=0,\theta_2=1)$ and $(\theta_1=1,\theta_2=0)$, where agents receive a penalty of $k$. To solve the coordination game, agents iteratively adjust their parameters in the direction of the value function gradients, i.e., $\sfrac{V}{\theta_1}$ and $\sfrac{V}{\theta_2}$ in na\"ive update rule, $\sfrac{V^{\text{LOLA}}}{\theta_1}$ and $\sfrac{V^{\text{LOLA}}}{\theta_2}$ in LOLA, and $\sfrac{V^{\text{LA}}}{\theta_1}$ and $\sfrac{V^{\text{LA}}}{\theta_2}$ in LA. Therefore, we can analyze the dynamics of $\theta_1$ and $\theta_2$ in LA, LOLA, and na\"ive agents to investigate their behaviors.
\begin{figure}
	\centering
	\vspace{\intextsep}
	\includegraphics[width=0.48\textwidth]{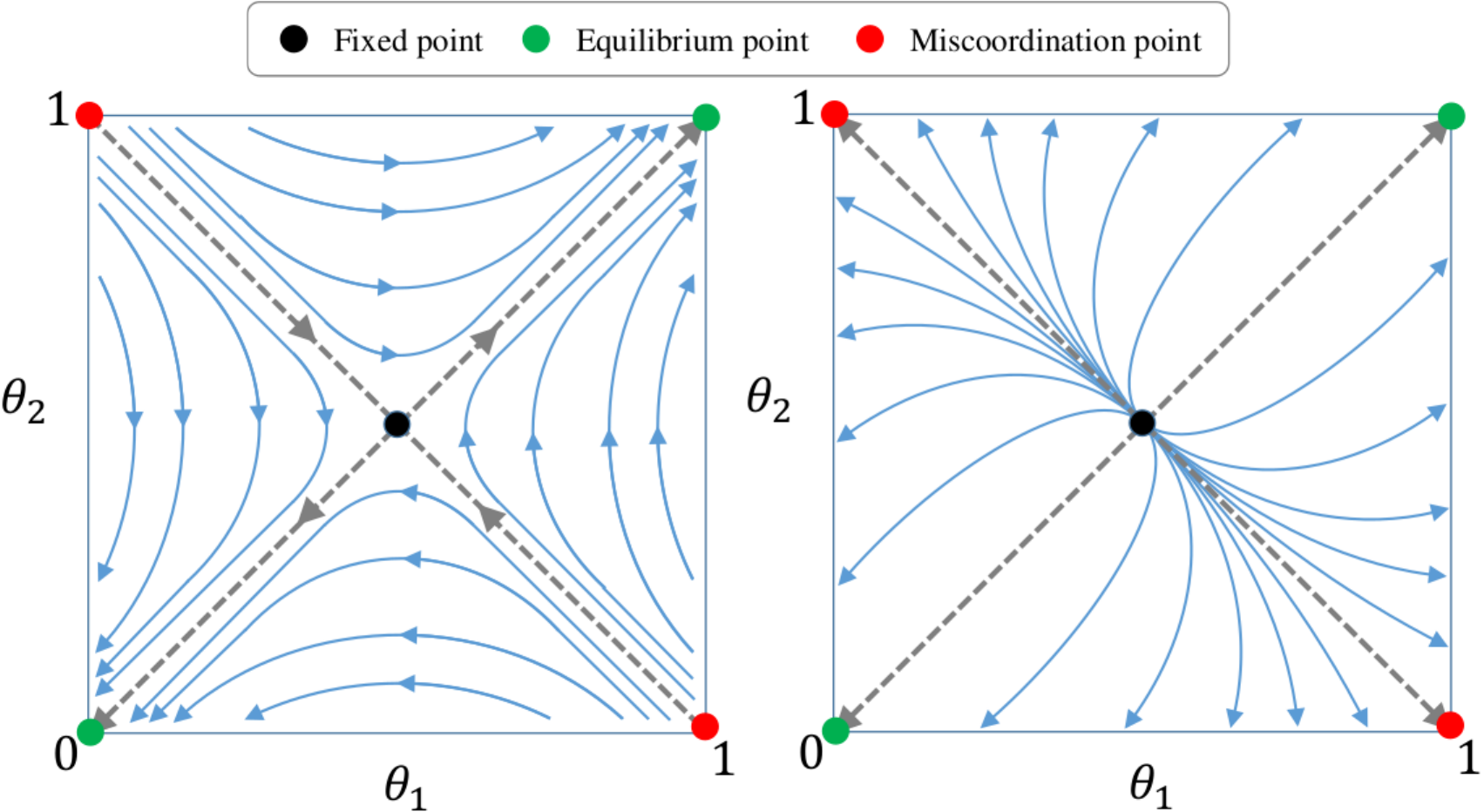}
	\caption{Phase planes of the LA and LOLA dynamics. Left: unstable saddle fixed point. Right: unstable fixed point.}
	\label{fig:phaseplane}
\end{figure}
\begin{theorem}
\label{th:misscoordination1}
    If, in the previously defined two-agent two-action coordination game with a miscoordination regret $g$, the agents are updated following the LA method and a fixed prediction length $\eta$, then they can be subject to miscoordination for $g>\tfrac{1}{2\eta}$. 
\end{theorem}
{\bf Proof}. Given Eq. \ref{eq:la}, the unconstrained dynamics of LA agents can be defined by the following differential equations:
\begin{equation}
\begin{split}
    \begin{bmatrix}\sfrac{d\theta_1}{d t}\\\sfrac{d\theta_2}{d t}\end{bmatrix} = \begin{bmatrix}4\eta g^2 & 2g\\2g & 4\eta g^2\end{bmatrix} \begin{bmatrix}\theta_1\\\theta_2\end{bmatrix} - \begin{bmatrix}2\eta g^2 + g\\2\eta g^2 + g\end{bmatrix}.
\end{split}
\end{equation}
This system of equations has a unique fixed point (zero gradients) at $\theta_1=\theta_2=0.5$ (see Figure \ref{fig:phaseplane}). The eigenvalue analysis of the coefficient matrix yields two real eigenvalues, $\lambda_1=4\eta g^2 + 2g$ and
$\lambda_2=4\eta g^2 - 2g$, and two respective diagonal and off-diagonal eigenvectors. While $\lambda_1$ is always positive, the sign of $\lambda_2$ depends on the values of both $\eta$ and $g$. For a fixed prediction length, non-positive values of $\lambda_2$ are reached by $g \leq \tfrac{1}{2\eta}$. In this case, the fixed point is an unstable saddle point (or unstable line in case of $\lambda_2=0$), and the agents, with any initial values of $\theta_1$ and $\theta_2$ (except on the fixed point itself), converge to the equilibrium points (see Figure \ref{fig:phaseplane}-Left). However, when the miscoordination regret increases, $g > \tfrac{1}{2\eta}$, the fixed point becomes an unstable (source) point (see Figure \ref{fig:phaseplane}-Right). Therefore, some initial values of $\theta_1$ and $\theta_2$ naturally lead to the miscoordination points $(\theta_1=0,\theta_2=1)$, and Theorem \ref{th:misscoordination1} is proved.

\begin{theorem}
\label{th:misscoordination2}
    If, in the previously defined two-agent two-action coordination game with a miscoordination regret $g$, the agents are updated following the LOLA method and a fixed prediction length $\eta$, then they can be subject to miscoordination for $g>\tfrac{1}{4\eta}$.
\end{theorem}
{\bf Proof}. Given the LOLA update rule in Eq. \ref{eq:lola}, the unconstrained dynamics can be defined as
\begin{equation}
\begin{split}
    \begin{bmatrix}\sfrac{d\theta_1}{d t}\\\sfrac{d\theta_2}{d t}\end{bmatrix} = \begin{bmatrix}8\eta g^2 & 2g\\2g & 8\eta g^2\end{bmatrix} \begin{bmatrix}\theta_1\\\theta_2\end{bmatrix} - \begin{bmatrix}4\eta g^2 + g\\4\eta g^2 + g\end{bmatrix}.
\end{split}
\end{equation}
This system of equations has a unique fixed point, again at $\theta_1=\theta_2=0.5$ (see Figure \ref{fig:phaseplane}). The eigenvalue analysis of the coefficient matrix yields two real eigenvalues, $\lambda_1=8\eta g^2 + 2g$ and
$\lambda_2=8\eta g^2 - 2g$, and two respective diagonal and off-diagonal eigenvectors. Similar to the case of LA agents, $\lambda_1$ is always positive, and the sign of $\lambda_2$ depends on the values of both $\eta$ and $g$. For a fixed prediction length, non-positive values of $\lambda_2$ are reached by $g \leq \tfrac{1}{4\eta}$. In this case, the fixed point is an unstable saddle point (or unstable line in case of $\lambda_2=0$), and the agents, with any initial values of $\theta_1$ and $\theta_2$ (except on the fixed point itself), converge to the equilibrium points. However, when the miscoordination regret increases, $g > \tfrac{1}{4\eta}$, the fixed point becomes an unstable (source) point. Therefore, some initial values of $\theta_1$ and $\theta_2$ naturally lead to the miscoordination points, and Theorem \ref{th:misscoordination2} is proved.

\begin{theorem}
\label{th:misscoordination3}
    If, in the previously defined two-agent two-action coordination game with a miscoordination regret $g$, the agents follow the na\"ive updates, then they are never subject to miscoordination for any value of $g$.
\end{theorem}
{\bf Proof}. In the case of the na\"ive agents, we have
\begin{equation}
\begin{split}
    \begin{bmatrix}\sfrac{d\theta_1}{d t}\\\sfrac{d\theta_2}{d t}\end{bmatrix} = \begin{bmatrix}0 & 2g\\2g & 0\end{bmatrix} \begin{bmatrix}\theta_1\\\theta_2\end{bmatrix} - \begin{bmatrix} g\\ g\end{bmatrix}.
\end{split}
\end{equation}
Similar to the case of LOLA and LA, this system of equations has a unique fixed point (zero gradients) at $\theta_1=\theta_2=0.5$. The eigenvalue analysis of the coefficient matrix yields two real eigenvalues, $\lambda_1=2g$ and
$\lambda_2=-2g$, and two respective diagonal and off-diagonal eigenvectors. This time, however, the eigenvalues are of opposite signs for any values of $g$, and the fixed point is always an unstable saddle point. Therefore, any initial values of $\theta_1$ and $\theta_2$ (except on the fixed point itself) naturally lead to the equilibrium points.  

A closer inspection of LOLA and LA methods reveals two important aspects of their fundamental ideas.
\begin{enumerate}
    \item Anticipating other agents' learning is only effective when it is close to their true future learning. Both LOLA and LA assume a reasoning order for other agents. If this assumption is wrong, it can negatively affect the coordination among the cooperative agents. For self-interested agents, it is natural for them to not unveil their true reasoning orders to each other as they have different goals. However, common-interested agents can benefit more from this reasoning information to achieve their common goal.
    \item The idea of shaping other agents' learning can be misleading if the other agents do not follow, making agents more likely to suffer from miscoordination. LOLA agents constantly underestimate each other, and each agent intends to shape the learning of others. Letcher et al. \cite{letcher2018stable} indicated that these arrogant behaviors lead to outcomes that are strictly worse for all agents. It is also clear from Theorems \ref{th:misscoordination1} and \ref{th:misscoordination2} that the range of $g$ that can lead to miscoordination in LOLA ($g>\tfrac{1}{4\eta}$) is larger than the range of $g$ in LA ($g>\tfrac{1}{2\eta}$).
\end{enumerate}
Given the above discussion, we hypothesize that if agents be informed of the reasoning orders and properly follow the shaping plans, they can improve coordination among themselves.


\section{Hierarchical Learning Anticipation}
In this section, we propose \textit{Hierarchical Learning Anticipation} (HLA), a methodology designed to improve coordination among fully-cooperative agents. In contrast to LOLA and LA, HLA determines a hierarchy among the agents to specify their reasoning orders. Specifically, we first assign $n$ agents to $n$ hierarchy levels, where levels one and $n$ are the lowest and highest hierarchy levels, respectively. In each hierarchy level, the assigned agent is a \textit{leader} of the lower hierarchy levels and a \textit{follower} of the higher ones, with two reasoning rules:
\begin{enumerate}
    \item A leader knows the reasoning orders of the followers and is one level higher
    \item A follower cannot shape the leaders and only follows their shaping plans
\end{enumerate}
With these reasoning rules, we can address the two previously mentioned shortcomings of LOLA and LA. Specifically, the first reasoning rule makes sure that a leader has correct assumptions about the followers' reasoning orders and, consequently, can accurately anticipate their future behaviors. With the second reasoning rule, we can control the shaping plans of the agents and make sure that the shaping plans are followed. Below, we describe the update rules in our proposed HLA.

\begin{algorithm}[t]
\begin{algorithmic}
    \STATE Initialize $\theta_i$ $\forall i \in \mathcal{N}$, the prediction length $\eta$, and the learning rate $\lambda$
    \STATE Randomly assign the agents into $n$ hierarchy levels.
    \STATE Rename the agent assigned to level $i$ as the agent $i$, $\forall i \in \mathcal{N}$
    \FOR{$\textrm{iter}=1\textrm{ to MaxIter}$}
        \FOR{level $i=n\textrm{ to }1$}
            \FOR{level $j=1\textrm{ to }i$}
            \STATE \textbf{if} $\;\;\,j=1 \;\&\; i\neq n$ \textbf{then} $\Delta \theta_j=\;\;\;\;\;\;\;\;\;\;\;\;\;\;\;\;\;\;\;\;\;\;\;\mathbb{E}_{\rho(s)}\eta\frac{\partial}{\partial \theta_j}V(s,\theta_1,...,\theta_{i},\bar{\theta}_{i+1},...,\bar{\theta}_n)$ 
            \STATE \textbf{elif} $j\neq 1 \;\&\; i=n$ \textbf{then} $\Delta \theta_j=\;\;\mathbb{E}_{\rho(s)}\eta\frac{\partial}{\partial \theta_j}V(s,\theta_1+\Delta \theta_1,...,\theta_{j-1}+\Delta \theta_{j-1},\theta_j,...,\theta_n)$ 
            \STATE \textbf{elif} $j=1 \;\&\; i= n$ \textbf{then} $\Delta \theta_j=\;\;\;\;\;\;\;\;\;\;\;\;\;\;\;\;\;\;\;\;\;\;\;\;\;\;\;\;\;\;\;\;\;\;\;\;\;\;\;\;\;\;\;\;\;\;\mathbb{E}_{\rho(s)}\eta\frac{\partial}{\partial \theta_j}V(s,\theta_1,...,\theta_n)$ 
            \STATE \textbf{else} $\Delta \theta_j=\;\;\;\;\;\;\;\;\;\;\;\;\;\;\;\;\;\;\;\;\;\;\;\;\;\;\;\;\;\;\;\;\;\;\;\;\;\;\;\;\;\;\;\;\;\;\mathbb{E}_{\rho(s)}\eta\frac{\partial}{\partial \theta_j}V(s,\theta_1+\Delta \theta_1,...,\theta_{j-1}+\;\;\;\;\;\;\;\;\Delta \theta_{j-1},\theta_j,\bar{\theta}_{j+1},...,\bar{\theta}_n)$
            \ENDFOR 
        \STATE Set $\bar{\theta}_i=\theta_i+\Delta \theta_i$
        \ENDFOR  
        \STATE Update $\theta_i=\theta_i + \lambda\frac{\Delta\theta_i}{\eta}$ $\forall i \in \mathcal{N}$  
    \ENDFOR
\end{algorithmic}
 \caption{HLA for a set of $n$ common-interested agents ($\mathcal{N}$).   }
 \label{alg:HLA}
\end{algorithm}

For simplicity, we set $n=2$, and we assume that agents one ($\theta_1$) and two ($\theta_2$) are assigned to the hierarchy levels one and two, respectively. In other words, agent one is a na\"ive follower, and agent two is a first-order leader. Based on our first reasoning rule, the leader performs first-order reasoning, and its gradient adjustment is similar to a first-order LOLA agent:
\begin{equation}
\label{eq:hr_leader}
\begin{split}
    \nabla_{\theta_2}V^{\text{HLA-Leader}}(s,\theta_1+\Delta\theta_1,\theta_2) \approx &\nabla_{\theta_2}V+ (\nabla_{\theta_1\theta_2}V)^\intercal \Delta\theta_1\\
    &+ (\nabla_{\theta_2} \Delta\theta_1)^\intercal \nabla_{\theta_1}V,
\end{split}
\end{equation}
where $V = V(s,\theta_1,\theta_2)$ is the common value function, and $\Delta\theta_1 =  \mathbb{E}_{\rho(s)}\eta\nabla_{\theta_1}V$. However, unlike first-order LOLA agents, the first-order leader knows the reasoning level of the follower, which is a na\"ive agent. The shaping plan of the first-order leader is to change its parameters as
\begin{equation}
\begin{split}
    \bar{\theta}_2=\theta_2 + \mathbb{E}_{\rho(s)}\eta\nabla_{\theta_2}V^{\text{HLA-Leader}}(s,\theta_1+\Delta\theta_1,\theta_2),
\end{split}
\end{equation}
so that an optimal increase in the common value is achieved after its new parameters are taken into account by the na\"ive follower. Therefore, based on our second reasoning rule, the na\"ive follower must follow the plan and adjust its parameters through 
\begin{equation}
\label{eq:hr_follower}
\begin{split}
    \nabla_{\theta_1}&V^{\text{HLA-Follower}}(s,\theta_1,\bar{\theta}_2)  \approx \nabla_{\theta_1}V\\
    &+ (\nabla_{\theta_2\theta_1}V)^\intercal \mathbb{E}_{\rho(s)}\eta\nabla_{\theta_2}V^{\text{HLA-Leader}}(s,\theta_1+\Delta\theta_2,\theta_2),
\end{split}
\end{equation}
Since the follower can no longer shape the leader, the shaping term in Eq. (\ref{eq:hr_follower}) is zero. Therefore, the gradient adjustment for the na\"ive follower in HLA is similar to a first-order LA agent, which predicts the leader parameters as $\theta_2 + \mathbb{E}_{\rho(s)}\eta\nabla_{\theta_2}V^{\text{HLA-Leader}}(s,\theta_1+\Delta\theta_1,\theta_2)$.

Algorithm \ref{alg:HLA} illustrates the HLA update rules for the case of $n$ agents. In the theorem below, we show how HLA agents can effectively avoid miscoordination in the two-agent two-action coordination game.

\begin{theorem}
\label{th:misscoordination_hr}
    If, in the previously defined two-agent two-action coordination game with a miscoordination regret $g$, the agents are updated following HLA, then they are not subject to miscoordination for any value of $g$.
\end{theorem}
{\bf Proof}. Given Eqs. \ref{eq:hr_leader} and \ref{eq:hr_follower}, the unconstrained dynamics of the HLA agents can be defined by the following differential equations 
\begin{equation}
\begin{split}
    \begin{bmatrix}\sfrac{\partial\theta_1}{\partial t}\\\sfrac{\partial\theta_2}{\partial t}\end{bmatrix} = \begin{bmatrix}  4\eta g^2&2g+16\eta^2g^3\\  2g&8\eta g^2\end{bmatrix} \begin{bmatrix}\theta_1\\\theta_2\end{bmatrix} - \begin{bmatrix}8\eta^2 g^3 + 4\eta g^2 + g\\4\eta g^2 + g\end{bmatrix},
\end{split}
\end{equation}
resulting in a unique fixed point at $\theta_1=\theta_2=0.5$ and two real eigenvalues, $\lambda=6\eta g^2 \pm 2p\sqrt{9\eta^2g^2+1}$. Unlike the case of LOLA and LA, the eigenvalues are now of opposite signs for any values of $g$, and the fixed point is always an unstable saddle point. Therefore, any initial values of $\theta_1$ and $\theta_2$ (except on the fixed point itself) naturally lead to the equilibrium points, and Theorem \ref{th:misscoordination_hr} is proved.

With Theorem \ref{th:misscoordination_hr}, we have shown that HLA naturally avoids miscoordination, and therefore, our hypothesis is correct. However, the main goal is to demonstrate that HLA improves coordination among common-interested agents with respect to na\"ive learning, which does not take into account learning anticipation at all. If HLA does not improve the coordination, there is no clear benefit over the na\"ive learners, as they also avoid miscoordination. 

\begin{figure}
	\centering
	\vspace{\intextsep}
	\includegraphics[width=0.48\textwidth]{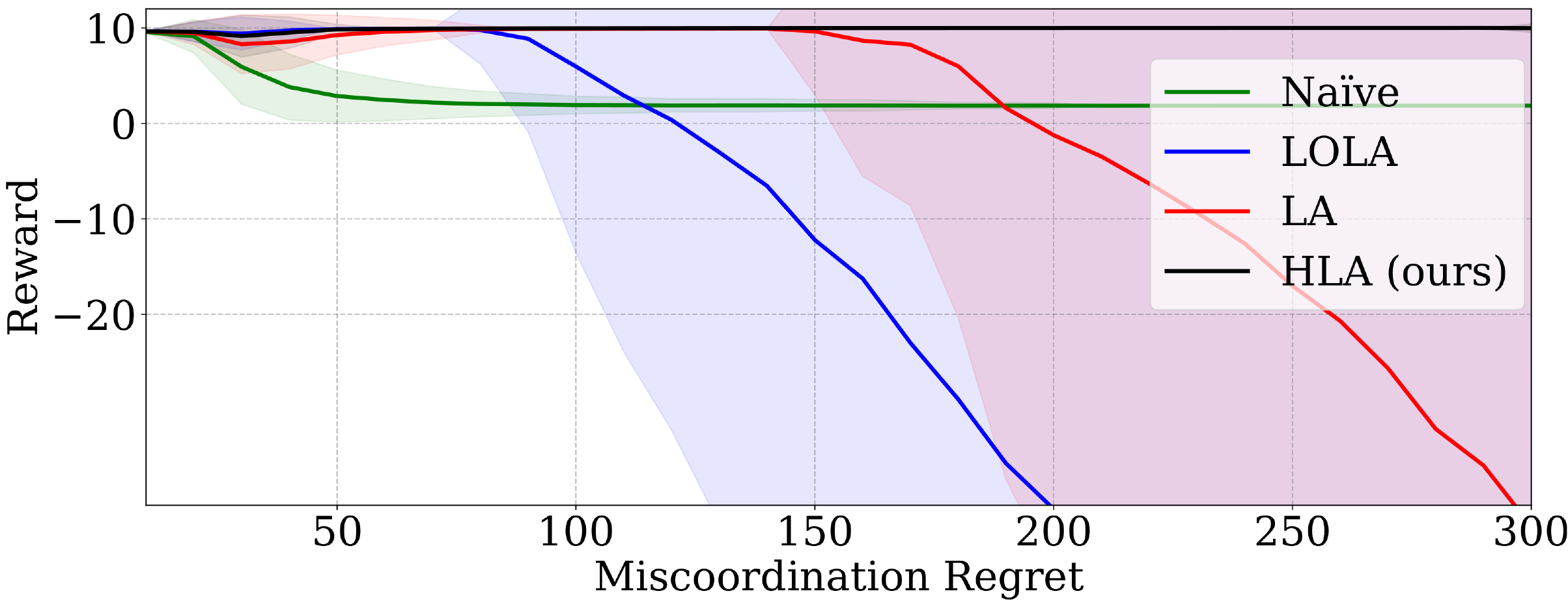}
	\caption{Converged results for various values of miscoordination regret in the three-action coordination game.}
	\label{fig:matrix_miscoord_3a}
\end{figure}

To further show the benefits of HLA, we employ a standard two-agent three-action coordination game \citep{claus1998dynamics}. The game has a common reward matrix 
\begin{equation}
\begin{split}
    \mathcal{R}_1=\mathcal{R}_2=\begin{bmatrix}10 & 0 & k\\0 & 2 & 0\\k & 0 & 10\end{bmatrix},
\end{split}
\end{equation}
and we define $g=10-k$ as the miscoordination regret. Each agent $i \in \{1,2\}$ is parameterized with three parameters: $\theta_i=\{\theta_i^1,\theta_i^2,\theta_i^3\}$ ($\theta_i^j>0\; \forall j \in \{1,2,3\}$ and $\sum_j\theta_i^j=1$), representing the probability of taking the actions one, two, and three, respectively. Consequently, the common value function of the game can be defined as
\begin{equation}
\begin{split}
    V_1=V_2=10(\theta_1^1\theta_2^1+\theta_1^3\theta_2^3)+2\theta_1^2\theta_2^2+k(\theta_1^1\theta_2^3+\theta_1^3\theta_2^1).
\end{split}
\end{equation}
The game has two global equilibrium points ($\theta_1^1=1,\theta_2^1=1$ and $\theta_1^3=1,\theta_2^3=1$), one local equilibrium point ($\theta_1^2=1,\theta_2^2=1$), and two miscoordination points ($\theta_1^1=1,\theta_2^3=1$ and $\theta_1^3=1,\theta_2^1=1$). 

In Figure \ref{fig:matrix_miscoord_3a}, we depict the converged results for this game for na\"ive, LA, LOLA, and HLA agents, for various values of miscoordination regret $g$. The experiments are run 500 times until convergence, with random initializations. For HLA, we randomly assigned the agents to the hierarchy levels (leader and follower) in each experiment. From Figure \ref{fig:matrix_miscoord_3a}, we find that both LA and LOLA agents are subject to miscoordination for high values of $g$, which is consistent with our findings for the two-action coordination game. However, the most interesting aspect of this experiment is that by increasing the value of $g$, the coordination among the na\"ive agents reduces, leading them to the local equilibrium point, whereas our HLA consistently achieves the highest reward, independently of the miscoordination regret. These results clearly show the benefits of HLA over other methods.

\section{Discussion and Future Work}
In this study, we extended the applicability of learning anticipation to games with fully-cooperative agents. We demonstrated that methods such as LOLA and LA, which heavily benefit from learning anticipation in general-sum games, can significantly reduce coordination when the agents are fully-cooperative. We first hypothesized that when agents know the reasoning orders of others and properly follow the shaping plans, they could improve coordination among themselves. To verify our hypothesis, we proposed the novel HLA method, which incorporates a hierarchy to regulate the reasoning orders and shaping plans of agents. Having HLA, we can now use the benefits of learning anticipation in fully-cooperative games. 

Nevertheless, there are still some unanswered questions about applying HLA to more complex fully-cooperative games. For instance, we limit ourselves to fully-cooperative normal-form games with differentiable objective functions where agents can access gradients and Hessians. In many multi-agent problems, the objective functions are non-differentiable. In these cases, the agents must estimate the higher-order gradients with various approximation methods. Furthermore, the agents may not have access to other agents' exact parameters, and they may need to infer the parameters from state-action trajectories. Consequently, future studies on the current topic are required.
\balance






\bibliographystyle{ACM-Reference-Format} 
\bibliography{sample}


\end{document}